\documentclass[pra,amsmath,aps,twocolumn]{revtex4}

\usepackage{graphicx}
\usepackage{rotating}

\newcommand{\bal}{\begin{align}}
\newcommand{\eal}{\end{align}}

\newcommand{\rhof}{\rho_{\rm f}^{}}
\newcommand{\rhofda}{\rho_{\rm f}^\dagger}
\newcommand{\Uf}{U_{\rm f}}

\newcommand{\be}{\begin{equation}} \newcommand{\ee}{\end{equation}}
\newcommand{\ba}{\begin{eqnarray}} \newcommand{\ea}{\end{eqnarray}}
\newcommand{\bastar}{\begin{eqnarray*}}
\newcommand{\eastar}{\end{eqnarray*}}

\newcommand{\TKK}{Laboratory of Physics, Helsinki University of Technology \\ P.~O.~Box 1100, FIN-02015 HUT, Finland}
\newcommand{\BERKELEY}{Department of Chemistry and Pitzer Center for Theoretical Chemistry,\\ University of California, Berkeley, CA 94720}
\begin{document}

\title{High fidelity one-qubit operations under random telegraph noise}

\author{Mikko~M\"ott\"onen}\email{mikko.mottonen@tkk.fi}\affiliation{\BERKELEY}\affiliation{\TKK}
\author{Rogerio~de~Sousa}\affiliation{\BERKELEY}
\author{Jun~Zhang}\affiliation{\BERKELEY}
\author{K.~Birgitta~Whaley}\affiliation{\BERKELEY}
\date{\today}

\begin{abstract}
We address the problem of implementing high fidelity one-qubit operations 
subject to time dependent noise in the qubit energy splitting. 
We show with explicit numerical results that high fidelity bit flips and
one-qubit {\footnotesize{NOT}} gates may be generated by imposing bounded 
control fields. For noise correlation times shorter than the time for a 
$\pi$-pulse, the time optimal $\pi$-pulse yields the highest fidelity. 
For very long correlation times, fidelity loss is approximately due to systematic error, 
which is efficiently tackled by compensation for off-resonance with a 
pulse sequence (CORPSE). For intermediate ranges of the noise correlation
time we find that short CORPSE, which is less accurate than 
CORPSE in correcting systematic errors, yields higher fidelities.
Numerical optimization of the 
pulse sequences using gradient ascent pulse engineering results in noticeable
improvement of the fidelities for the bit flip and marginal improvement
for the {\footnotesize{NOT}} gate.

\end{abstract}

\hspace{5mm}

\maketitle
\section{Introduction}\label{sec1}

In physical implementations of quantum computers, one of the most
challenging tasks is to find an efficient and experimentally feasible
way to overcome the problems caused by undesired interactions between the quantum bits, qubits, and their surrounding environment. These interactions, which destroy
the quantum interference between qubit states, lead to errors
and loss of fidelity, a phenomenon generally referred to as decoherence.


A variety of methods to fight decoherence
have been proposed in the literature, including 
error correcting codes~\cite{shor1995,steanne1996}, 
decoherence free subspace coding~\cite{zanardi1997,lidar1998},  
noiseless subsystem coding~\cite{knill2000},
dynamical decoupling~\cite{viola1998,viola1999,viola2003}, 
quantum feedback control~\cite{wiseman1994a,wiseman1994b}, and
quantum reservoir engineering~\cite{poyatos1996}.
Most of these schemes are not efficient in making full use of all the physical resources. 
For example, encoding schemes employing decoherence free subspaces or error
correcting codes store the quantum information in a specific
portion of the whole qubit space, or encode several physical qubits
into one logical qubit. 
Design and applicability of such codes depends on the nature of
the decoherence sources and imposes additional requirements on encoding
and decoding.
Dynamical decoupling schemes possess the attractive feature that they require no ancillary qubits, since the interactions between the qubits and the environment are effectively canceled out by applying external control fields. 
It has also been shown that such decoupling can be realized
using finite energy soft pulses~\cite{viola2003}, and that it is possible
to carry out qubit rotations without disturbing the decoupling
process~\cite{viola1999}.  However, dynamical decoupling is based on
stroboscopic pulsing of the qubit, at a rate significantly
faster than the usual characteristic frequency of environmental
fluctuations. This requires strong control fields, which
might cause technical problems in the laboratory. For
example, it has been pointed out that the high energy deposition
needed for dynamical decoupling of nuclear spins is incompatible with
the low temperature requirement in some qubit
implementations~\cite{ladd2005}.

In this paper, we consider the design of fidelity optimized
one-qubit operations in a noisy environment. Motivated by 
experiments in solid state qubits~\cite{nakamura3,collin2004}, we
assume random telegraph noise (RTN)~\cite{vankampen} in the qubit
energy splitting as a phenomenological model for the environmental
fluctuations. Whereas an ensemble of RTN fluctuators models the
ubiquitous $1/f$ noise in electronic circuits~\cite{kogan}, solid
state devices on the nanoscale are often found to be affected by a
single RTN source~\cite{fujisawa2000, kurdak1997} characterized by its correlation time $\tau_{\rm c}$. 
Although we assume that the noise couples only to the qubit energy splitting, 
a similar analysis can be made for errors in the rotation angle~\cite{brown2004,morton2005}.

Composite pulses are known to provide an efficient way to reduce errors due to systematic off-resonant perturbations, e.g., compensation for off-resonance
with a pulse sequence (CORPSE)~\cite{cummins2000}.  Here we focus on the
situation in which the perturbation of a qubit is fluctuating in time, and seek to suppress the decoherence arising from this time-dependent noise by imposition of a bounded control field.
We are particularly interested in the regime where the maximal 
energy $a_{\rm max}$ provided by the control field is related to the 
noise
correlation time as $\tau_{\rm c}a_{\rm max}/\hbar\sim 1$, since this is an 
important regime for the experimentalists where dynamical decoupling is not 
applicable. In this regime, we find that a novel optimized pulse profile can
increase the fidelity of quantum operations by up to 30\% in comparison
to the standard composite pulse sequences such as CORPSE and short CORPSE
(SCORPSE)~\cite{cummins2000,cummins2003}. 
However, we provide an analysis over the 
entire range of correlation times, ranging from short
noise correlation times satisfying $\tau_{\rm c}a_{\rm max}/\hbar\ll 1$,
where the optimal fidelity is obtained by pulsing at minimum time, to the
long correlation time regime, $\tau_{\rm c}a_{\rm max}/\hbar\gg 1$,
where to a good approximation, the fidelity loss is due to
systematic time-independent errors.  
The latter case is efficiently tackled by the composite pulse approach~\cite{cummins2000,cummins2003}.

It was stated in Ref.~\cite{cummins2003} that CORPSE is the shortest sequence in the family of composite pulse sequences correcting systematic errors as efficiently as possible and composing of up to three pulses. Therefore, it was considered to be the most useful one.  
Whereas SCORPSE is not as accurate as CORPSE, it is 
shorter in time. Thus it may still be of some interest, depending on the 
physical scenario. We illustrate this fact here by showing that SCORPSE 
actually yields higher fidelities than CORPSE in the regime of
intermediate noise correlation time, $\tau_{\rm c}a_{\rm max}/\hbar\sim 1$,
for the one-qubit example studied here.  The optimal performance of CORPSE is
in fact limited to just the long correlation time regime. 
We also go beyond these composite pulse sequences to obtain fidelity
optimized pulses consisting of large numbers of pulse amplitudes that
are numerically derived using an adaptation of 
the method of gradient ascent pulse engineering (GRAPE)~\cite{khaneja2005}.
Although GRAPE was originally developed for finding control pulses in closed quantum 
systems, it is utilized here to the determination of bounded control pulses in
an noisy quantum system. We find that numerical optimization of a pulse
sequence with GRAPE noticeably increases the fidelity of bit flip operations,
compared with the 
results of the standard composite pulse sequences. In contrast to this improvement for a bit flip, 
only a rather marginal improvement in fidelity is
found on employing GRAPE to numerically generate a fidelity-optimized
control pulse sequence for the complete one-qubit {\footnotesize{NOT}} gate.

The remainder of this paper is organized as follows. In Sec.~\ref{sec2}, we
characterize the system Hamiltonian, the noise model, and the
fidelity. Section~\ref{sec3} introduces the pulse sequence generation
methods we use for the noisy qubit system, and Sec.~\ref{sec4} presents
the results obtained for implementation of a state transformation corresponding
to a bit flip and a quantum gate corresponding to the one-qubit 
{\footnotesize{NOT}} gate.  Finally, Sec.~\ref{sec6} concludes the paper
with a discussion of extensions and possible generalizations.

\section{System characterization}\label{sec2}
We consider a single qubit described by the effective Hamiltonian 
\be
H=\sum_{i\in\{x,y,z\}} [a_i(t)+\eta_i(t)]\sigma_i,
\ee
where the symbols $\{\sigma_i\}$ denote the Pauli spin
matrices~\cite{NielsenChuang}, $\{\eta_i(t)\}$ are the amplitudes of the 
environmental noise, and $\{a_i(t)\}$ are the external control fields.
Note that the latter are parameterized here by their corresponding energy 
amplitudes, rather than the actual physical control fields, e.g., electric field amplitudes.  We assume
that the strength of the control fields is finite, and denote their
maximum possible value by $a_{\rm max}$. To simplify the discussion, we 
consider only the case when there is no
control in the $y$ or $z$ directions, and no noise in the $x$ and $y$ 
directions. Under these assumptions, the Hamiltonian becomes 
\be\label{dhami} 
H=a(t)\sigma_x+\eta(t)\sigma_z,
\ee 
where the control field $a:=a_x\in [-a_{\rm max},a_{\rm max}]$ and we have 
used the notation $\eta(t):=\eta_z(t)$.

For RTN, the amplitude of the noise $\eta$ changes randomly in time between two values $-\Delta$ and $\Delta$. The quantity $\Delta$ describes the strength of the noise, and the frequency of the jumps  between $-\Delta$ and $\Delta$ is determined by the correlation time $\tau_{\rm c}$. Specifically, the probability of the noise to jump in an infinitesimal time interval $dt$ is given by $\frac{dt}{\tau_{\rm c}}$. Hence, the probability of no jumps taking place in a time interval of length $t$ is
\be\label{pn}
p_0(t)=e^{-t/\tau_{\rm c}}.
\ee
In generating sample trajectories of RTN, Eq.~(\ref{pn}) can also be inverted 
to yield the sojourn time before a jump takes place. Thus we get a sample 
trajectory of RTN by taking random numbers $p_i\in(0,1)$ and then deriving the corresponding jump time instants
\be\label{ti}
t_i=\sum_{j=1}^i-\tau_{\rm c}\,{\rm ln}(p_j).
\ee
Using the values of these jump times, we can express the noise process $\eta(t)$ as
\be
\eta(t)=(-1)^{\sum_i\Theta(t-t_i)}\eta(0),
\ee
where $\Theta(t)$ is the Heaviside step function.

Since we use an effective Hamiltonian operating solely on the qubit rather than treating the full quantum dynamics of both the qubit and the environment, 
we need to average over different noise trajectories in order
to obtain the system dynamics under the influence of RTN. Therefore, the dynamics of the system density matrix $\rho$ can be written as
\be\label{eq:9}
\rho(t)=\lim_{N\to\infty}\frac{1}{N}\sum_{k=1}^NU_k\rho_0U_k^\dagger,
\ee
where $\rho_0=\rho(0)$ is the initial state of the system and the operators 
$\{U_k\}$ refer to unitary time evolution of the system under a certain trajectory
$\eta_k(t)$. Formally, the operator $\{U_k\}$ is written as
\be
U_k=\mathcal{T}e^{-i\int_0^t d\tau \left[a(\tau)\sigma_x+\eta_k(\tau)\sigma_z\right]/\hbar},
\ee
where $\mathcal{T}$ is the time ordering operator.

Let $\rhof$ be the desired final state of the system. Following Ref.~\cite{khaneja2005}, we define the fidelity function as
\be\label{fid}
\phi(\rhof,\rho_0)={\rm tr}\{\rhofda\rho(T)\},
\ee
where $\rho(T)$ is the actual state of the system at the final time instant $T$.
Substituting Eq.~(\ref{eq:9}) into Eq.~(\ref{fid}), we obtain
\be\label{eq:10}
\phi(\rhof,\rho_0)=\lim_{N\to\infty}\frac{1}{N}\sum_{k=1}^N{\rm tr}\{\rhofda U_k\rho_0U_k^\dagger\}.
\ee
Equation~\eqref{eq:10} shows that the fidelity function defined here can be viewed as an average over fidelity functions corresponding to individual 
unitary time developments in noiseless quantum systems that are characterized
by the evolution operators $U_k$.

Let us write the initial state of the qubit as
\be
\rho_0=(I+c_x\sigma_x+c_y\sigma_y+c_z\sigma_z)/2,
\ee
where $c_i$ are real numbers.
For implementation of quantum gates rather than state transformations, we 
define a fidelity function as the average over all pure initial conditions
of the qubit:
\be\label{fidin}
\Phi(\Uf)=\frac{1}{4\pi}\int_{c_x^2+c_y^2+c_z^2=1} {\rm d}\Omega\,\phi(\Uf\rho_0\Uf^\dagger,\rho_0),
\ee
where the operator that we desire to implement is denoted by $\Uf$ and ${\rm d}\Omega$ is an infinitesimal solid angle on the Bloch sphere. Simplification of the integral in Eq.~(\ref{fidin}) yields
\be
\Phi(\Uf)=\frac{1}{2}+\lim_{N\to\infty}\frac{1}{12N}\sum_{k=1}^N\sum_{j=1}^3{\rm tr}\{\Uf\sigma_j\Uf^\dagger U_k\sigma_jU_k^\dagger\}.
\ee

\section{Pulse sequences for noisy systems}\label{sec3}
In this section, we introduce the pulse sequences that will be used for suppression of decoherence. We first summarize the two composite pulse sequences CORPSE and SCORPSE~\cite{cummins2000,cummins2003} that were originally designed to 
correct systematic errors in the implementation of one-qubit quantum gates.  
The control fields corresponding to the CORPSE pulse sequence are
\be\label{corpse} 
a_{\rm C}(t)=\left\{
\begin{array}{rll}
a_{\rm max}, & {\rm for} & 0<t'< \pi/3 \\
-a_{\rm max}, & {\rm for} & \pi/3\le t'\le 2\pi \\
a_{\rm max}, & {\rm for} & 2\pi<t'< 13\pi/3,
\end{array}\right.  
\ee 
where $t$ is related to the dimensionless time $t'$ by $t'=a_{\rm max}t/\hbar$.
For the SCORPSE pulse sequence we have the control fields 
\be\label{scorpse} 
a_{\rm SC}(t)=\left\{
\begin{array}{rll}
-a_{\rm max}, & {\rm for} & 0<t'< \pi/3 \\
a_{\rm max}, & {\rm for} & \pi/3\le t'\le 2\pi \\
-a_{\rm max}, & {\rm for} & 2\pi<t'< 7\pi/3.
\end{array}\right.  
\ee 
In the absence of noise, the CORPSE and SCORPSE pulse sequences generate both the one-qubit {\footnotesize{NOT}} gate and the bit flip state transformation.

An alternative to these composite pulse sequences is provided by numerical
construction of pulse sequences optimized for maximum fidelity.  
Such fidelity-optimized sequences may be constructed by an adaptation of the
GRAPE algorithm~\cite{khaneja2005} which was originally designed to steer the 
dynamics of coupled nuclear spins. No noise effects or bounds on control fields are included in the
original implementation.  For full details of the GRAPE algorithm
for closed quantum systems, see Ref.~\cite{khaneja2005}. 

The key feature of the GRAPE algorithm is to approximate the continuous pulse shape on a time interval $[0, T]$ by a function that is constant on $n$ small 
time intervals of length $\Delta t=T/n$, and then to derive the corresponding gradients 
of the fidelity function with respect to these constant values. Let $U_k^m$ be 
the unitary time evolution operator corresponding to the time interval 
$[(m-1)\Delta t, m\Delta t]$ and to the noise trajectory $\eta_k$. 
In this interval, the control field is approximated by a constant, $a^m$. 
Since the fidelity function $\phi(\rhof,\rho_0)$ is an average of the 
fidelity functions used in Ref.~\cite{khaneja2005}, the gradient of 
$\phi(\rhof,\rho_0)$ is obtained as an average of the gradients derived in  Ref.~\cite{khaneja2005}. Thus
\be\label{grad}
\frac{\delta\phi(\rhof,\rho_0)}{\delta a^m}=-\frac{i\Delta t}{\hbar}\lim_{N\to\infty}\frac{1}{N}\sum_{k=1}^N{\rm tr}\{(\lambda^m_k)^\dagger[\sigma_x,\rho^m_k]\},
\ee
where 
\be
\lambda^m_k=(U^n_kU^{n-1}_k\cdots U^{m+1}_k)^\dagger\rhof U^n_kU^{n-1}_k\cdots U^{m+1}_k,
\ee
and 
\be
\rho^m_k=U^m_kU^{m-1}_k\cdots U^1_k\rho_0(U^m_kU^{m-1}_k\cdots U^1_k)^\dagger.
\ee
In the case that there exist other control terms 
$\{a_{k}^{\rm c}(t)H^{\rm c}_k\}$ in the Hamiltonian, the corresponding 
gradients can be obtained from Eq.~(\ref{grad}) by substituting $a$ by 
$a_{k}^{\rm c}$ and $\sigma_x$ by $H^{\rm c}_k$.

We note that for $\Delta=0$, all the individual RTN trajectories are identical 
and consequently the averaging and limiting procedures in Eq.~(\ref{grad}) can 
be omitted. In this case, Eq.~(\ref{grad}) reduces to the equation for 
noiseless systems derived in Ref.~\cite{khaneja2005}.

To derive the gradient of the average fidelity $\Phi$, we note that
\be\label{vt}
\Phi(\Uf)=\frac{1}{2}+\frac{1}{12}\sum_{j=1}^{3}\phi
(\Uf\sigma_j\Uf^\dagger,\sigma_j).
\ee 
Hence, the gradient of Eq.~\eqref{vt} can be obtained from
Eq.~(\ref{grad}) as 
\be\label{gradin} 
\frac{\delta\Phi(\Uf)}{\delta
  a^m}=\frac{1}{12}\sum_{j=1}^3
\frac{\delta\phi(\Uf\sigma_j\Uf^\dagger,\sigma_j)}{\delta a^m}.  
\ee

In the GRAPE algorithm, we calculate the gradient of the desired fidelity function using Eq.~(\ref{grad}) or~(\ref{gradin}), and update the control fields by moving along the direction of the gradient with the restriction $a\in [-a_{\rm max}, a_{\rm max}]$.  
This procedure results in an optimized pulse sequence for a given operation 
time $T$. Moreover, the fidelity is also optimized with respect to the operation time.

We note that the pulse sequences yielding the optimal fidelity for each set of system parameters are not unique. 
In order to find as smooth and as simple sequence as possible, we therefore
start from a constant control field and use the gradient method to maximize 
the fidelity. To ascertain whether our solution 
achieves a local or the global maximum in fidelity, we repeated the procedure
for several different, uncorrelated initial values of the control field.  This
resulted in different pulse sequences with equal fidelities, suggesting 
that we have indeed found the global maximum, though this cannot be 
conclusively claimed. Thus, when we refer to the results of the GRAPE 
algorithm, we shall describe the corresponding pulse sequences as optimized 
rather than optimal.

\section{High fidelity one-qubit operations}\label{sec4}
In this section, we present optimized pulse sequences implementing high 
fidelity one-qubit operations that were obtained using the GRAPE algorithm, 
and compare the results with those from the standard CORPSE and SCORPSE 
pulse sequences.  We restrict out attention here to two quantum operations
on the one-qubit system, namely, the state transformation corresponding to
a bit flip, and the one-qubit NOT gate.

\subsection{Bit flip}\label{bitflip}
We consider a one-qubit bit flip, i.e., flipping a one-qubit state from one of 
the two poles of the Bloch sphere to the other. This problem may arise, 
for example, when some qubits of a multi-qubit register need to be flipped 
to reach a non-trivial state after a collective initialization. The initial 
and final states can be taken as
\be\label{inc}
\rho_0=
\begin{pmatrix}
0 & 0 \\ 0 & 1
\end{pmatrix}\quad {\rm and} \quad
\rhof=
\begin{pmatrix}
1 & 0 \\ 0 & 0
\end{pmatrix}.
\ee

We first consider the limiting cases: vanishing noise correlation time, 
and infinite noise correlation time.

\noindent{\emph{Case 1:}} $\tau_{\rm c}\to 0$. 
In this case, RTN averages out due to the well-known phenomenon of motional 
narrowing~\cite{slichter} since the noise changes its sign so rapidly that 
there is no time for the qubit to drift into the direction of the noise at any
given time. It is therefore not surprising that a time optimal $\pi$-pulse 
\be\label{pip} 
a_\pi(t)=a_{\rm max},\quad {\rm for}\quad t\in[0, \pi\hbar/a_{\rm max}],
\ee 
is also fidelity optimal, since it is impossible to correct arbitrary fast switching using bounded controls.

\noindent{\emph{Case 2:}} $\tau_{\rm c}\to \infty$. In this limiting case,  
RTN reduces 
to a constant drift, whereas for large but finite $\tau_c$ the drift may be
treated as approximately constant.  In comparison to a $\pi$-pulse, pulse sequences such as CORPSE and
SCORPSE that are specifically designed to correct 
systematic errors will clearly improve the fidelity of the desired quantum operation here.  

\begin{figure}[tbh]
\includegraphics[width=200pt]{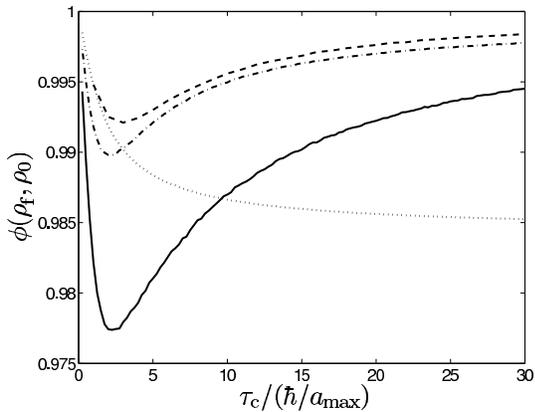}
\caption{\label{fig1} Fidelities $\phi(\rhof,\rho_0)$ as functions of the noise
correlation time $\tau_{\rm c}$ for a $\pi$-pulse (Eq.~(\ref{pip}), 
dotted line), CORPSE (Eq.~(\ref{corpse}), solid line), and SCORPSE (Eq.~(\ref{scorpse}), dash-dotted line). 
The optimized fidelity found using GRAPE is shown as the dashed line. 
The strength of the RTN in this example is $\Delta=0.125\times a_{\rm max}$.}
\end{figure}

In Fig.~\ref{fig1}, the fidelities obtained from 
$\pi$-pulse, CORPSE, SCORPSE, and GRAPE pulse sequences for a bit flip are plotted as 
functions of the correlation time $\tau_{\rm c}$. The noise strength 
$\Delta$ is chosen to be $0.125\times a_{\rm max}$ in this example.  
As expected, GRAPE yields the highest fidelities for all values of noise correlation time $\tau_c$ since it enforces optimization of the pulse sequence.  Note that the fidelity curve of GRAPE has a global minimum near the
correlation time $\tau_{\rm c}\approx 3\hbar/a_{\rm max}$.  The existence of the minimum is due to
the fact that since the GRAPE pulse sequences are optimized, they will not only yield perfect unit fidelity in the short correlation time limit 
$\tau_{\rm c}\to 0$ as a result of motional narrowing, but they will also
yield unit fidelity in the long correlation time limit $\tau_{\rm c}\to\infty$.
The latter argument is true since small systematic errors can be corrected to arbitrary 
accuracy~\cite{Geen1991} and provided that the GRAPE algorithm does find
the global optimal solution.
Consequently, there must be a minimum at a finite value of $\tau_c$ in the 
fidelity curve generated by GRAPE.  The corresponding 
fidelity curves of the CORPSE and the SCORPSE pulse 
sequences also show such minima which result from
the fact that these sequences are specifically designed to correct the
systematic errors.
In contrast, the fidelity curve for the $\pi$-pulse is seen to be a 
monotonically decreasing function of the correlation time, reflecting the fact
that this pulse cannot correct systematic errors. 
Figure~\ref{fig1} also provides a good example of the general result 
that for intermediate noise correlation times SCORPSE is more favorable than 
CORPSE.  This is a consequence of the shorter operation time of SCORPSE, which
is more significant at finite values of $\tau_c$ than the fact that CORPSE is
more efficient than SCORPSE in correcting systematic errors. This
balance between the length of the operation time and the accuracy in correcting systematic errors 
results in a cross-over between the fidelity curves of SCORPSE and CORPSE at
very large correlation times. Hence, the CORPSE curve eventually rises above the
SCORPSE curve for longer $\tau_{\rm c}$ than shown in Fig.~\ref{fig1}.

\begin{figure}[tbh]
\includegraphics[width=200pt]{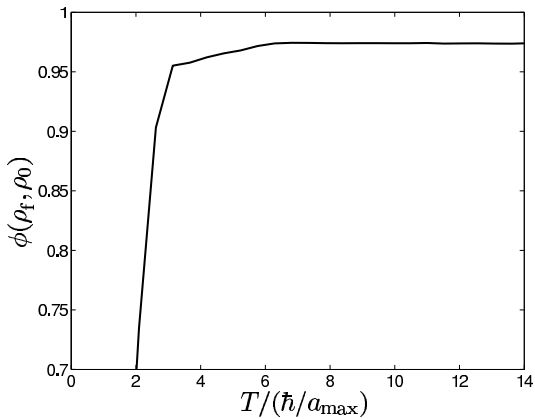}
\caption{\label{fig2} Fidelity $\phi(\rhof,\rho_0)$ of the GRAPE pulse 
sequence as a function of the operation time $T$, for noise correlation time 
$\tau_{\rm c}=5\hbar/a_{\rm max}$ and noise strength 
$\Delta=0.25\times a_{\rm max}$. 
}
\end{figure}

As discussed in Sec.~\ref{sec3}, the GRAPE algorithm for finding
optimal pulse sequences involves optimization of the operation
time $T$. However, the RTN used in this work does not have any dynamical
effect on the initial density matrix $\rho_0$. Moreover, the noise is 
Markovian and hence any pulse $a'(t)$ with operation time $T'<T$ may be 
extended to an operation time $T$ without change of fidelity by setting 
\be
\label{monotone}
a(t)=\left\{
\begin{array}{ccc}
0, & {\rm for} & 0<t<T-T' \\
a'(t-T+T'), & {\rm for} & T-T'<t<T.
\end{array}
\right.
\ee
Thus the fidelity is a monotonically increasing function of 
the operation time $T$. In fact, it is found that the fidelity saturates at a 
maximum value for rather short operation times, see for example 
Fig.~\ref{fig2}. 

\begin{figure}[tbh]
\includegraphics[width=200pt]{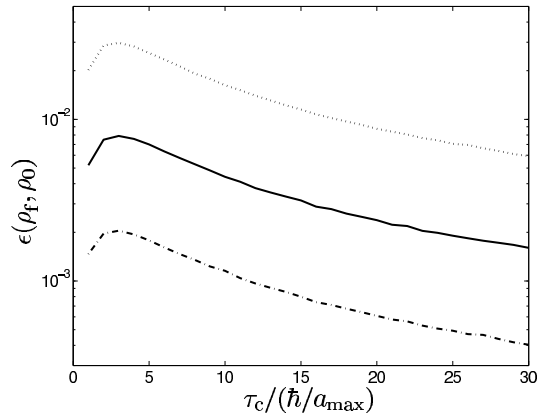}
\caption{\label{fig3} Error $\epsilon(\rhof,\rho_0)=1-\phi(\rhof,\rho_0)$ 
of the GRAPE optimized pulse sequence, shown as a function of the correlation 
time $\tau_{\rm c}$ for noise strengths $\Delta=0.25\times a_{\rm max}$ 
(dotted line), $\Delta=0.125\times a_{\rm max}$ (solid line), and 
$\Delta=0.0625\times a_{\rm max}$ (dash-dotted line). }
\end{figure}

Figure~\ref{fig3} shows the error $\epsilon(\rhof,\rho_0)=1-\phi(\rhof,\rho_0)$ 
of the GRAPE optimized result as a function of the correlation 
time $\tau_{\rm c}$, for several different noise strengths. A quadratic 
dependence of the error on the noise strength, $\epsilon\propto \Delta^2$, is 
observed over the parameter ranges $\Delta\in[a_{\rm max}/16, a_{\rm max}/4]$ and $\tau_{\rm c}\in[0, 30\hbar/a_{\rm max}]$.

\subsection{One-qubit {\footnotesize{NOT}} gate}
In this subsection, we analyze implementations of one-qubit {\footnotesize{NOT}} gates under
RTN. The {\footnotesize{NOT}} gate corresponds to a $\pi$-rotation about the $x$-axis on the Bloch sphere and thus also carries out the bit flip considered in Sec.~\ref{bitflip}. For quantum gate implementations, we use the fidelity function $\Phi$ defined in Eq.~(\ref{fidin}).

\begin{figure}[tbh]
\includegraphics[width=200pt]{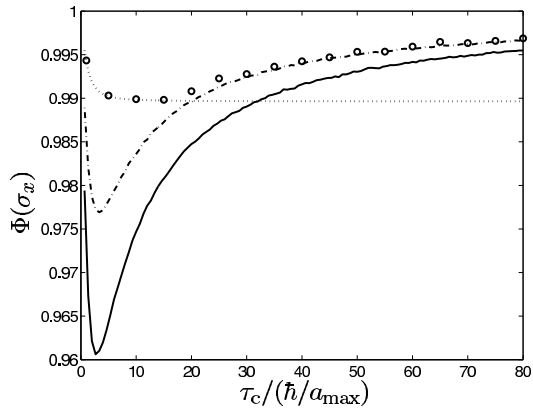}
\caption{\label{fig4} Fidelities $\Phi(\sigma_x)$ for the {\footnotesize{NOT}}
gate, shown as functions of the correlation time $\tau_{\rm c}$ for a 
$\pi$-pulse (dotted line), CORPSE (solid line), and SCORPSE (dash-dotted 
line). 
The figure also shows the optimized fidelity found using the GRAPE algorithm 
(circles). The RTN strength $\Delta$ is set at $0.125\times a_{\rm max}$.}
\end{figure}

The $\pi$-pulse, CORPSE, and SCORPSE sequences are specifically designed to 
implement a {\footnotesize{NOT}} gate. 
However, the GRAPE pulse sequences for a bit flip and a {\footnotesize{NOT}}
gate differ since the optimized fidelity functions are different for these two operations. 
Figure~\ref{fig4} shows the fidelities obtained with the $\pi$-pulse, CORPSE, 
SCORPSE, and GRAPE pulse sequences, as functions of the noise correlation time 
$\tau_{\rm c}$. The noise strength is set to be 
$\Delta=0.125\times a_{\rm max}$, as in Fig.~\ref{fig1}.  In comparison to the fidelities 
for the bit flip shown in Fig.~\ref{fig1}, the fidelities for the {\footnotesize{NOT}} gate derived 
under the CORPSE and SCORPSE pulse sequences are lower, whereas the fidelity under the 
$\pi$-pulse is higher. Nevertheless, Figs.~\ref{fig1} 
and~\ref{fig4} show qualitatively the same phenomena, namely, motional narrowing in the short
time correlation limit $\tau_{\rm c}\to 0$ and correction of systematic errors 
in the long time correlation limit $\tau_{\rm c}\to\infty$.

\begin{figure}[tbh]
\includegraphics[width=200pt]{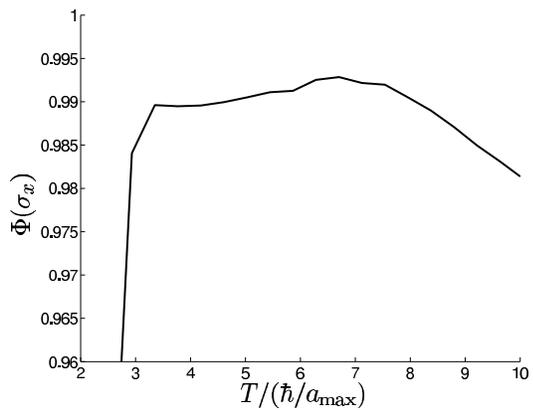}
\caption{\label{fig5} Optimized fidelity $\Phi(\sigma_x)$ obtained from
GRAPE, shown as a function of the operation time $T$ for noise correlation 
time $\tau_{\rm c}=30\hbar/a_{\rm max}$ and noise strength 
$\Delta=0.125\times a_{\rm max}$.}
\end{figure}

Since we employ the averaged gradient in Eq.~(\ref{gradin}) which effectively 
involves three gradients for fixed initial conditions, one might conclude that 
finding the optimized pulse sequences for quantum gates will require 
approximately three times as much computational time as for the bit flip. 
However, as noted above, the GRAPE algorithm finds the optimal operation time. 
This task is straightforward in the bit flip case, where as shown above, the fidelity is a monotonically increasing function of the operation time.
For the {\footnotesize{NOT}} gate however, the optimization is nontrivial. 
Because of the averaging over initial conditions, the optimal
fidelity does not necessarily increase monotonically with $T$, as illustrated in Fig.~\ref{fig5}.  
Finding the optimal 
operation time for a quantum gate thus clearly increases the complexity of the 
problem.

\begin{figure}[tbh]
\includegraphics[width=200pt]{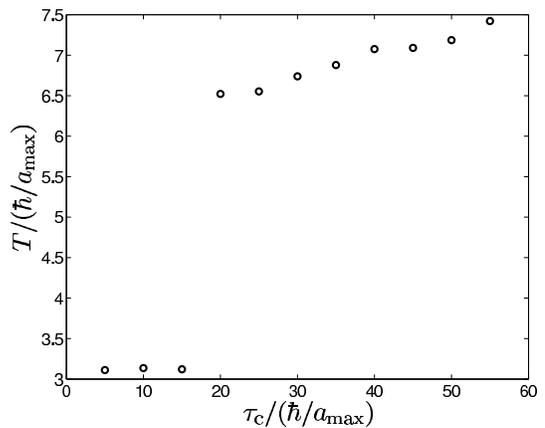}
\caption{\label{fig7} Optimal operation time of pulse sequences obtained
numerically with the GRAPE algorithm, shown as a function of the correlation 
time $\tau_{\rm c}$ for noise strength $\Delta=0.125\times a_{\rm max}$. }
\end{figure}

Additional insight into the efficiency of the GRAPE pulse sequence may be 
obtained by examining the behavior of the optimal operation time as a function 
of the correlation time. As shown in Fig.~\ref{fig7}, the optimal operation 
time of GRAPE increases sharply and approaches the time of the SCORPSE pulse 
sequence at a value $\tilde{\tau}_{\rm c}\approx 18\hbar/a_{\rm max}$.  The
resulting fidelity also becomes very close to that obtained with SCORPSE, see Fig.~\ref{fig4}.  
It appears from Fig.~\ref{fig4} that errors due to RTN 
cannot be efficiently corrected with bounded controls for correlation times shorter than $\tilde{\tau}_{\rm c}$, and therefore
the optimal operation time of GRAPE reduces to that of a $\pi$-pulse.

\section{Conclusions}\label{sec6}
In this work, we have studied how to perform high fidelity quantum operations 
on a one-qubit system that is subject to random telegraph noise acting on the 
qubit energy splitting. We considered examples of two major types of quantum
operations, namely, a state transformation and a quantum gate.  For the state
transformation we chose a bit flip in which the one-qubit state is flipped 
from the south pole of the Bloch sphere to the north pole, whereas for the
quantum gate we used the one-qubit {\footnotesize{NOT}} gate which generates 
a complete $\pi$-rotation on the Bloch sphere about the $x$-axis.

In both cases, we compared the fidelities obtained with the standard 
$\pi$-pulse, CORPSE, and SCORPSE pulse sequences.  The same qualitative 
phenomena were obtained for all three types of pulse sequence. In the limit of vanishing correlation time, motional narrowing occurs and implies the $\pi$-pulse to be the most accurate sequence since it is time optimal in 
implementing the {\footnotesize{NOT}} gate or a bit flip. 
On the other hand, the CORPSE sequence yields the highest fidelity in the long 
correlation time limit $\tau_{\rm c}\to \infty$, since it is designed to 
efficiently correct systematic errors.  Over a rather wide intermediate range of the correlation time $\tau_{\rm c}$, SCORPSE yields the highest
 fidelity among all these three pulse sequences, suggesting that it may be a useful approach to
 suppress environmental noise in physical realizations of
 quantum computers.

Furthermore, we obtained fidelity optimized pulse sequences using the GRAPE algorithm which was always found to yield higher fidelities than the most accurate composite pulse sequence.
Especially in the bit flip case, GRAPE yields noticeably higher fidelities than the composite pulse sequences.  In contrast, GRAPE introduces only a rather marginal improvement over the most accurate composite pulse sequence for the implementation of the complete {\footnotesize{NOT}} gate.

The results of this paper provide useful bounds for the implementation of high fidelity one-qubit operations in a noisy system without ancillary qubits. 
Although a simple RTN model is used in this paper, we expect that the qualitative dependency of the fidelity on noise strength and correlation time will also
be present in a general qubit system.  To investigate the validity of this conjecture, one may apply the methods presented here to the study of different noise models, e.g., Gaussian noise with a $1/f$ spectrum. 

Other extensions of this work include high fidelity control of multi-qubit systems. Recent work has addressed optimal control of noiseless coupled super conducting qubits~\cite{spoerl2005}.
For these systems, the environmental noise may act on
each qubit in either a correlated or uncorrelated fashion, which together with the entanglement of the qubits, expands the spectrum of the studies. 
It is an important open question to find a control sequence for the inter-qubit coupling term that implements a 
controlled {\footnotesize{NOT}} gate with high fidelity in the presence of noise. To generalize the methods of this paper to noisy multi-qubit systems, a reformulation of the equations for the fidelity and its gradient is required.

\begin{acknowledgments}
We thank the NSF for financial support under ITR Grant No. EIA-0205641, and DARPA and ONR under Grant No. FDN0014-01-1-0826 of the DARPA SPINs program. MM acknowledges the Academy of Finland, the Finnish Cultural Foundation, and Jenny and Antti Wihuri's foundation for financial support. 
We would like to express our
appreciation to S.~J.~Glaser and V.~Shende for helpful discussions.
\end{acknowledgments}

\bibliographystyle{prsty}
\bibliography{eqis05}

\end{document}